\documentclass{elsart}

\usepackage{graphicx}
\usepackage{subfigure}
\usepackage{amssymb}
\usepackage{color}

\begin{document}

\begin{frontmatter}

\title{Magnetic ordering in trigonal chain compounds}

\author[l1]{V.\ Eyert\corauthref{corauth}},
\corauth[corauth]{Corresponding author. fax: +49 821 598 3262}
\ead{eyert@physik.uni-augsburg.de}
\author[l1]{U.\ Schwingenschl\"ogl},
\author[l1]{C.\ Hackenberger},
\author[l1]{T.\ Kopp},
\author[l2]{R.\ Fr\'esard}, \and 
\author[l1]{U.\ Eckern}

\address[l1]{Institut f\"ur Physik, Universit\"at Augsburg, 
             86135 Augsburg, Germany}
\address[l2]{Laboratoire Crismat, UMR CNRS-ENSICAEN (ISMRA) 6508, 
             Caen, France}

\begin{abstract}
We present electronic structure calculations for the one-dimensional 
magnetic chain compounds $ {\rm Ca_3CoRhO_6} $ and $ {\rm Ca_3FeRhO_6} $. 
The calculations are based on density functional theory and the local 
density approximation. We use 
the augmented spherical wave (ASW) method. The observed alternation of 
low- and high-spin states along the Co-Rh and Fe-Rh chains is related 
to differences in the oxygen coordination of the transition metal sites.
Due to strong hybridization the O $ 2p $ states are polarized, giving 
rise to extended localized magnetic moments centered at the high-spin 
sites. Strong metal-metal overlap along the chains leads to a substantial 
contribution of the low-spin Rh $ 4d_{3z^2-r^2} $ orbitals to the 
exchange coupling of the extended moments. Interestingly, this mechanism 
holds for both compounds, even though the coupling is ferromagnetic for 
the cobalt and antiferromagnetic for the iron compound. However, our 
results allow to understand the different types of coupling from the 
filling dependence of the electronic properties. 
\end{abstract}

\begin{keyword}
density functional theory \sep 
low-dimensional compounds \sep 
magnetic chains \sep
geometrical frustration
\PACS 71.20.-b \sep 75.10.Pq \sep 75.30.Et
\end{keyword}
\end{frontmatter}

\section{Introduction}

The perovskite family of oxides shows a great compositional variety. 
Recently, a new structural class of hexagonal perovskites has started 
to attract increasing attention due to their extraordinary magnetic 
properties. These materials are based on the general formula
$ {\rm A'_3ABO_6} $ ($ {\rm A' =} $ alkaline earth; $ {\rm A, B =} $
transition metal) and crystallize in the trigonal $ {\rm K_4CdCl_6} $
structure \cite{stitzer01} with space group $ R\bar{3}c $ ($ D_{3d}^{6} $,
No.\ 167). In these compounds, transition metal-oxygen polyhedra form
well separated chains running parallel to the trigonal axis; see 
Refs.\ \cite{fjellvag96,aasland97} for a representation of the crystal
structure. Space between
the chains is filled with the $ {\rm A'} $ cations. Each chain consists of
alternating, face-sharing $ {\rm AO_6} $ trigonal prisms and $ {\rm BO_6} $
octahedra. Typically, the ratio of interchain to small intrachain metal-metal
distance is of the order of two, this fact explaining the importance of 
metal-metal bonding and the pronounced
one-dimensionality. The chains themselves are arranged on a triangular
lattice. As a consequence, in addition to showing the abovementioned unique
properties these compounds allow to study geometric frustration effects
such as partial disorder and spin-glass like behaviour.

A prominent member of this class is $ {\rm Ca_3Co_2O_6} $, 
which has two inequivalent cobalt sites, one in octahedral environment
(labelled Co1) and the other (Co2) centered in the trigonal prisms
\cite{fjellvag96,aasland97}. Neutron diffraction measurements revealed 
low- and high-spin moments of $ 0.08 \mu_B $ and $ 3.00 \mu_B $, 
respectively, at these sites. Together with susceptibility measurements 
they point towards a ferromagnetic coupling along the chains, and an 
antiferromagnetic one in the buckling a-b plane. Susceptibility 
\cite{kageyama97,hardy03a}, specific heat \cite{hardy03a}, and 
$ \mu $-SR \cite{sugiyama05} measurements clearly demonstrate the onset 
of three-dimensional ordering at $ {\rm T_{C_1}} = 24 $\,K, while a 
second transition was found at $ {\rm T_{C_2}} = 12 $\,K 
\cite{kageyama97,maignan00}. Magnetization measurements showed a 
plateau at $ \approx 1.31 \mu_B $ per f.u.\ for low field and a 
steep increase by a factor of three at about 3.5\,T, which was 
interpreted as a ferri- to ferromagnetic transition
\cite{aasland97,kageyama97}. 
In addition, the magnetization versus field curves display a rich 
structure below 12\,K \cite{hardy04}. In particular, the magnetization 
at saturation exceeds the (spin only) expected value of 4 $\mu_B$/f.u., 
indicating that the angular momentum is not fully quenched.
Electronic structure calculations revealed i) low- and high-spin moments, 
respectively, at the octahedral and trigonal prismatic cobalt sites in 
good agreement with the experimental data, ii) a rather large oxygen 
moment due to polarization by the high-spin cobalt sites, which together 
give rise to the formation of extended but still well localized 
$ {\rm CoO_6} $ moments, and iii) strong contributions 
to the ferromagnetic intrachain coupling from the $ 3d $ states of the 
low-spin octahedral cobalt atoms \cite{eyert04}. Recently, mapping
these results onto a Heisenberg model we were able to underline the
importance of the interplay of different crystal field splittings,
$ d $-$ p $ hybridizations, and metal-metal overlap for the ferromagnetic
intrachain order \cite{laschinger04,fresard04}. 

Nevertheless, the origin of the ferromagnetic intrachain coupling is 
still a matter of intense debate. In order to resolve the issue we 
have turned to the related compounds $ {\rm Ca_3CoRhO_6} $ and 
$ {\rm Ca_3FeRhO_6} $, which have the same crystal structure as 
$ {\rm Ca_3Co_2O_6} $ as well as very similar electronic properties 
\cite{dissudo}. However, while $ {\rm Ca_3CoRhO_6} $ likewise shows 
ferromagnetic intrachain order, the coupling is antiferromagnetic 
in the iron compound. Hence, by studying these systems we will 
contribute to a deeper understanding of the underlying exchange 
mechanisms. 

To be specific, according to susceptibility and neutron diffraction 
measurements $ {\rm Ca_3CoRhO_6} $ undergoes two transitions at 
$ {\rm T_{C1}} = 90 $\,K and $ {\rm T_{C2}} = 30 $\,K, which are 
attributed to the ferromagnetic intrachain and the antiferromagnetic 
interchain coupling, respectively \cite{niitaka99,niitaka01,niitaka02,davis03}. 
However, the specific heat did not exhibit any strong anomaly at these 
temperatures \cite{hardy03b}. Thus, elucidating the nature of the 
transitions requires more investigations.
Again, the interaction across the chains is antiferromagnetic driving 
the system into a partially disordered antiferromagnetic (PDA) state 
at intermediate temperatures. However, application of small magnetic 
fields results in a ferrimagnetic phase \cite{niitaka01b}. Neutron 
diffraction data suggest both 
the cobalt and rhodium ions to be trivalent \cite{niitaka01,loewenhaupt03}. 
Spin states are $ S = 2 $ at the trigonal prismatic cobalt sites and 
$ S = 0 $ at the octahedral rhodium sites. The magnetic moment amounts 
to 3.7\,$\mu_B$ per cobalt ion and is oriented parallel to the $ c $-axis. 

The magnetic ordering of $ {\rm Ca_3FeRhO_6} $ turns out to be less 
complex than in the cobaltate. Susceptibility data reveal a single 
transition into a three-dimensional antiferromagnetic phase at 12\,K 
\cite{niitaka99,davis03} with the easy axis oriented parallel to the 
chains as well as divalent iron in a $ S = 2 $ high-spin state and 
tetravalent rhodium in a $ S=1/2 $ low-spin state. In contrast, 
M\"ossbauer experiments were interpreted in favour of trivalent iron 
and $ S=5/2 $ \cite{niitaka03}. For the saturation magnetization a 
value of 3.74\,$\mu_B$ per f.u.\ has been given \cite{niitaka03}.

\section{Methodology}

The density functional calculations were performed using the
scalar-relativistic augmented  spherical wave (ASW) method
\cite{wkg,revasw}. In order to represent the correct shape of
the crystal potential in the large voids of the open crystal
structure, additional augmentation spheres were inserted. Optimal
augmentation sphere positions as well as radii of all spheres were
automatically generated by the sphere geometry optimization (SGO)
algorithm \cite{eyert98b}. Self-consistency was achieved by an
efficient algorithm for convergence acceleration \cite{mixpap}.
Brillouin zone sampling was done using an increased number of
$ {\bf k} $-points ranging from 28 to 770 points within the
irreducible wedge.

\section{Results and Discussion}

\begin{figure}[htb]
\centering 
\subfigure{\includegraphics[width=0.48\textwidth,clip]{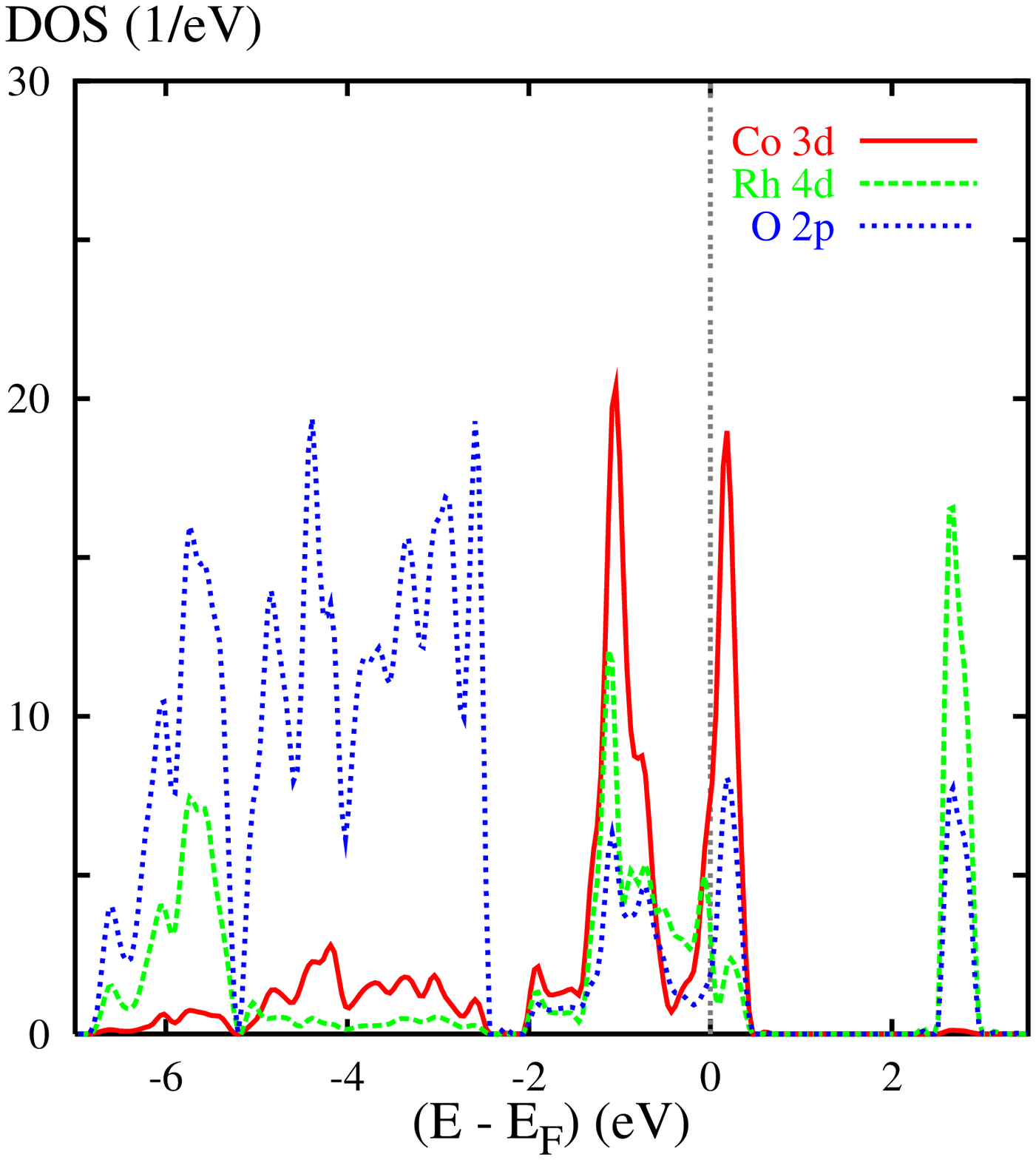}}
\subfigure{\includegraphics[width=0.48\textwidth,clip]{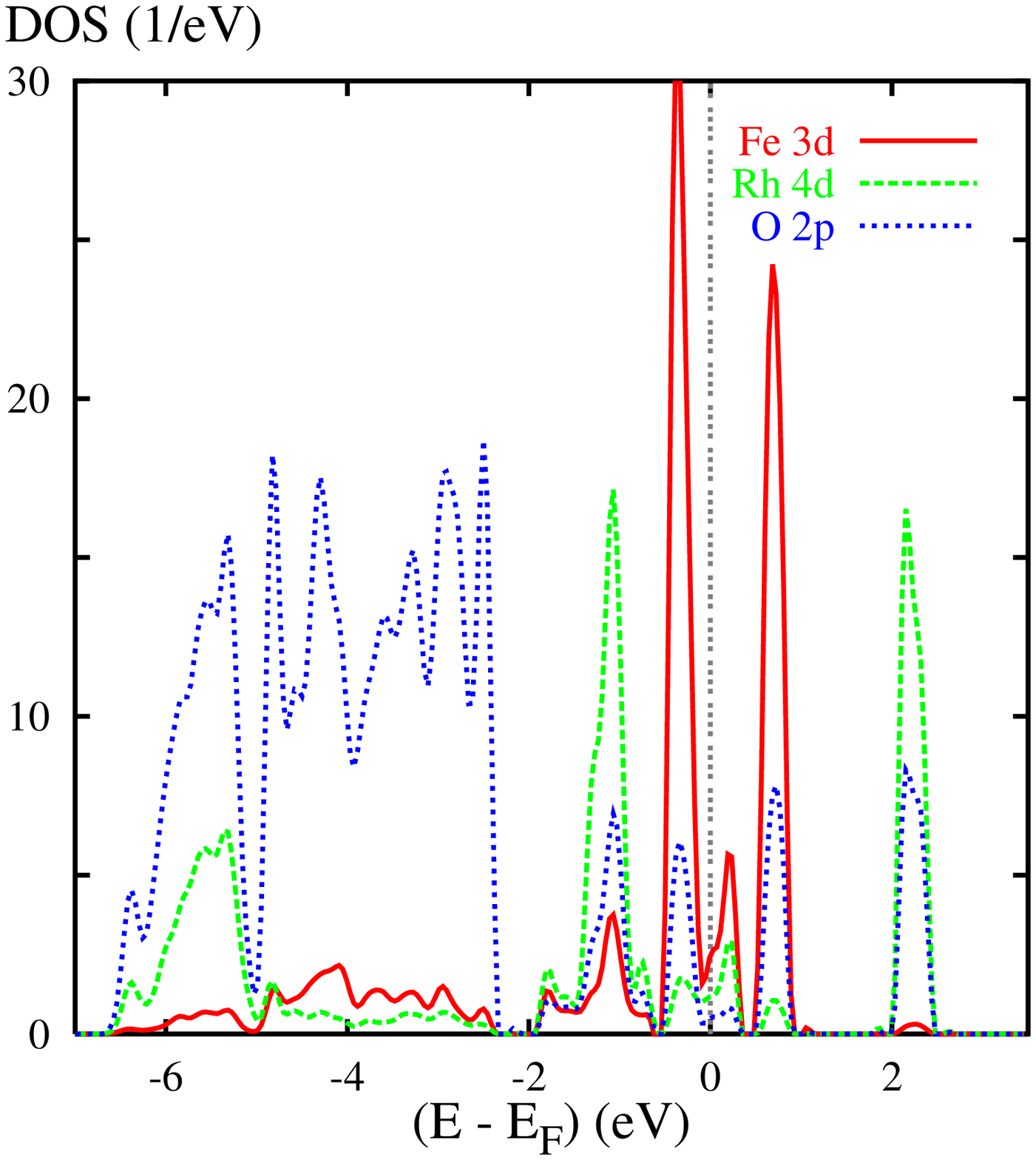}}
\caption{Partial densities of states of $ {\rm Ca_3CoRhO_6} $ 
         and $ {\rm Ca_3FeRhO_6} $.}
\label{fig2}
\end{figure}

Using the powder data of Niitaka {\em et al.} \cite{niitaka99}, we 
performed, in a first step, a set of calculations, where spin-degeneracy 
was enforced \cite{dissudo}. The results are shown in Fig.\ \ref{fig2}, 
which includes the partial densities of states of both $ {\rm Ca_3CoRhO_6} $ 
and $ {\rm Ca_3FeRhO_6} $. While we observe O $ 2p $ dominated bands between 
$ -6.8 $ and $ -2.4 $\,eV, three groups of bands of mainly metal 
$ d $-character are found at higher energies. Due to $ d $--$ p $ 
hybridization substantial $ p $/$ d $ contributions appear above/below 
$ -2 $\,eV, reaching up to 40\% in the upper valence and conduction bands. 

At the rhodium sites, the octahedral crystal field leads to nearly 
perfect splitting of the $ 4d $ states into occupied $ t_{2g} $ 
and empty $ e_g $ states. In contrast, the trigonal crystal field 
at the cobalt/iron sites results in a splitting into non-degenerate 
$ d_{3z^2-r^2} $ as well as doubly degenerate $ d_{xy,x^2-y^2} $ and 
$ d_{xz,yz} $ states. While, due to strong $ \sigma $-type $ d $--$ p $ 
bonding, the Rh $ 4d $ $ e_g $ states form the high-energy peak 
above $ 2.0 $\,eV, the peak below traces back to Co/Fe 
$ d_{xz,yz} $ states. As a consequence, spin-polarization of the 
latter bands is highly favourable this explaining the observed 
high-spin/low-spin scenario. 

Spin-polarized calculations performed in a second step lead to the 
observed ferromagnetic and antiferromagnetic behaviour found for 
$ {\rm Ca_3CoRhO_6} $ and $ {\rm Ca_3FeRhO_6} $, respectively. As 
a result, we obtain well localized magnetic moments of 
$ 0.48\,\mu_B $ (Rh), $ 2.59\,\mu_B $ (Co), $ 0.14\,\mu_B $ (O), 
and $ 0.00\,\mu_B $ (Ca) for the cobaltate as well as 
$ 0.00\,\mu_B $ (Rh), $ 3.72\,\mu_B $ (Fe), $ 0.14\,\mu_B $ (O), 
and $ 0.01\,\mu_B $ (Ca) for the ferrate. These values reflect the 
experimental result of low- and high-spin states at the octahedral 
and trigonal prismatic sites, respectively. The total moment per 
unit cell (i.e.\ per two formula units) amounts to $ 7.94\,\mu_B $ 
for $ {\rm Ca_3CoRhO_6} $ and $ \pm 4.59\,\mu_B $ per sublattice 
in $ {\rm Ca_3FeRhO_6} $. Worth mentioning are the rather high 
magnetic moments at the oxygen sites arising from the strong 
$ d $--$ p $ hybridization, which sum up to about $ 1\,\mu_B $ per 
trigonal prism. Adding to the $ 3d $ moment they lead to the formation 
of extended localized moments already observed in $ {\rm Ca_3Co_2O_6} $ 
\cite{eyert04} and confirm the formal Fe $ S=5/2 $ and Rh $ S=0 $ spin 
configurations. 

According to a more detailed analysis the increased iron moment as 
compared to that of cobalt arises from the energetical upshift of 
the bands due to the reduced electron count 
and can be attributed to the $ d_{xy,x^2-y^2} $ orbitals. However, 
since these orbitals point perpendicular to the chain axis their 
contribution affects mainly the size of the local moment and to a 
much lesser degree the intrachain coupling. 

\begin{figure}[htb]
\centering 
\includegraphics[width=0.95\textwidth]{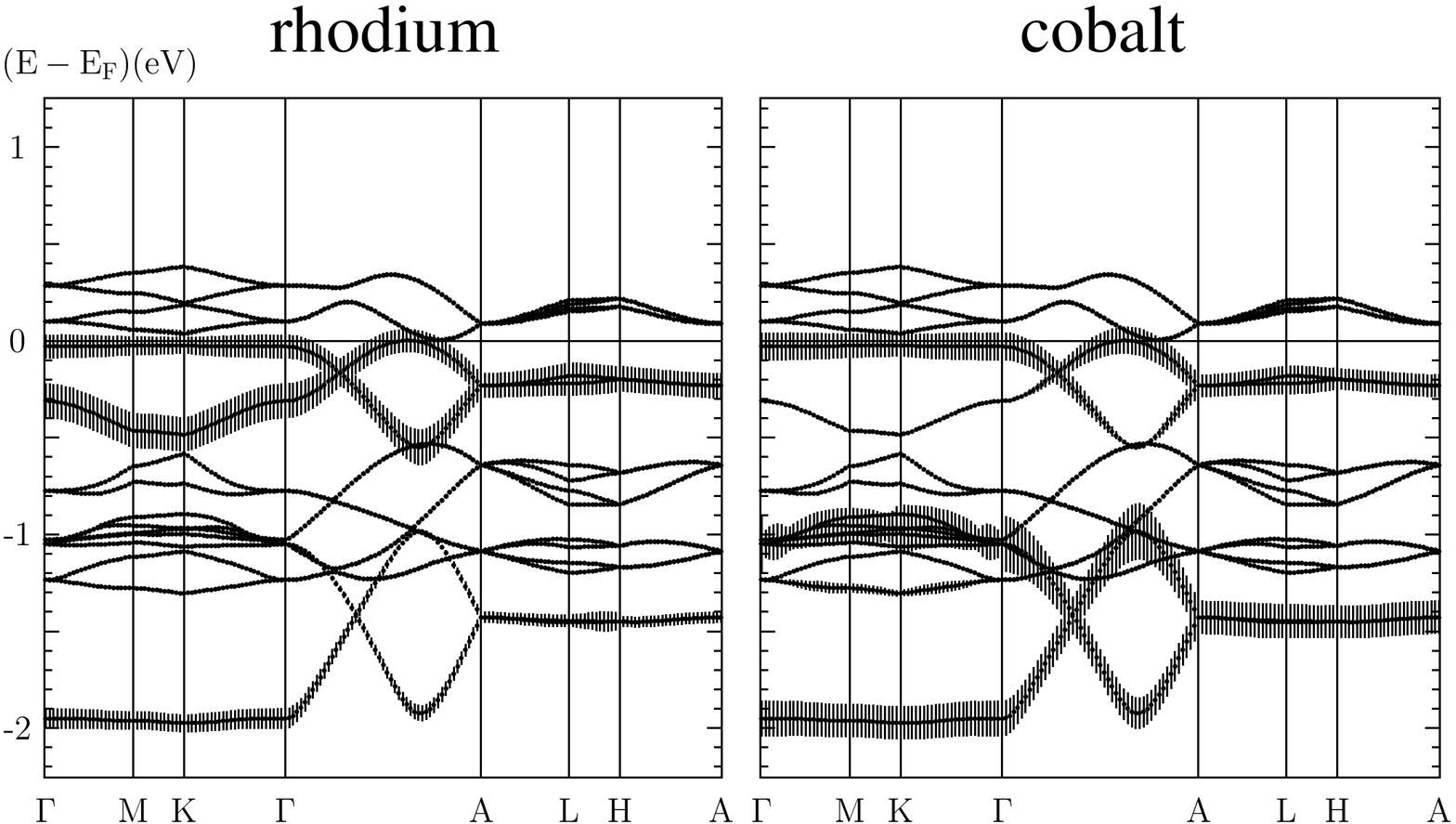}
\caption{Weighted electronic bands of spin-degenerate $ {\rm Ca_3CoRhO_6} $.
         Contributions of the Rh $ 4d_{3z^2-r^2} $ (left) and 
         Co $ 3d_{3z^2-r^2} $ (right) orbitals are indicated by bars.} 
\label{fig3}
\end{figure}

\begin{figure}[htb]
\centering
\includegraphics[width=0.95\textwidth,clip]{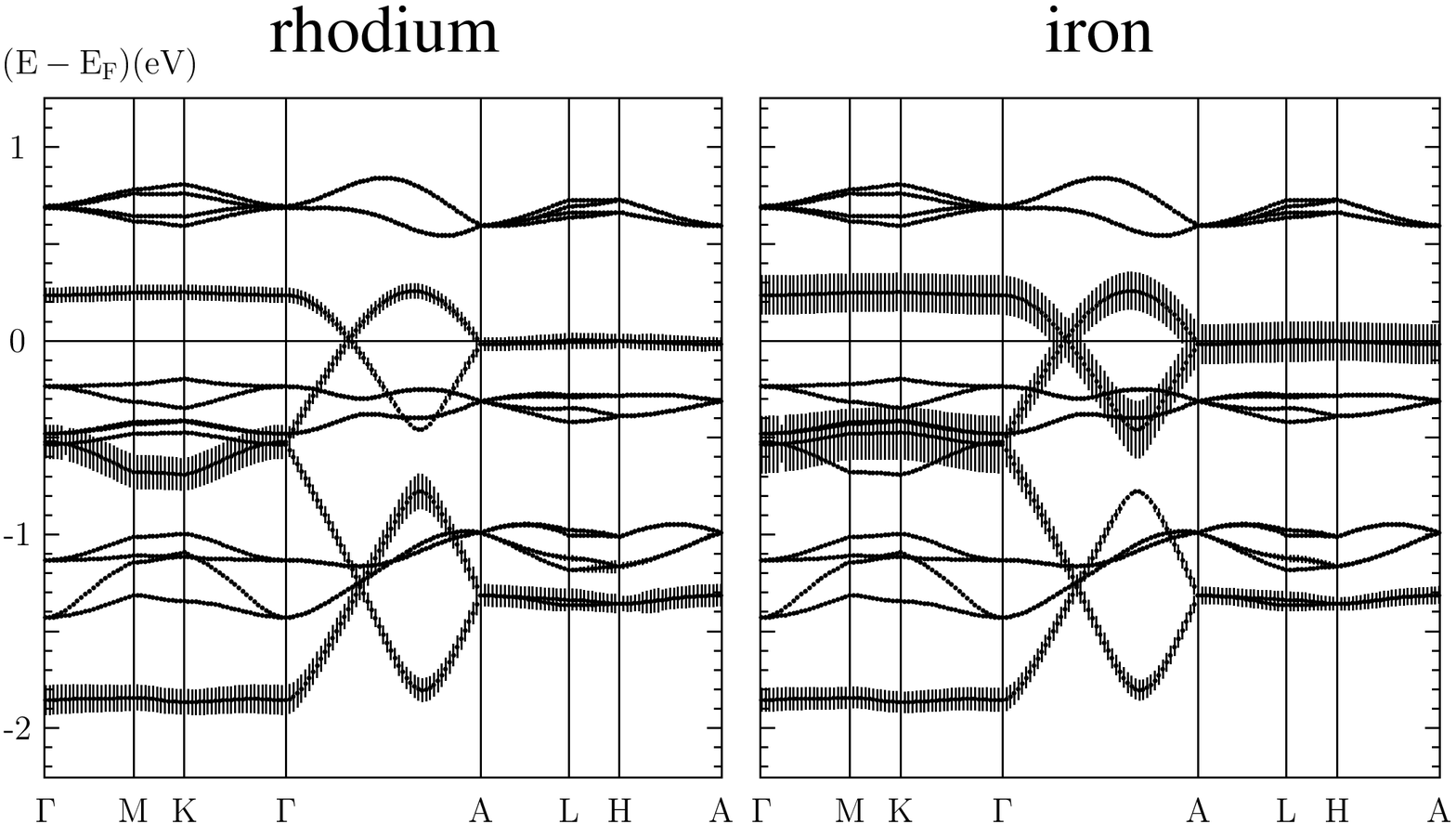}
\caption{Weighted electronic bands of spin-degenerate $ {\rm Ca_3FeRhO_6} $.
         Contributions of the Rh $ 4d_{3z^2-r^2} $ (left) and 
         Fe $ 3d_{3z^2-r^2} $ (right) orbitals are indicated by bars.}
\label{fig4}
\end{figure}

In order to understand this puzzling result we turn back to the 
spin-degenerate calculations and investigate the near-$ {\rm E_F} $ 
bands, which are displayed in Figs.\ \ref{fig3} 
and \ref{fig4}.  
The length of the bars appended to the bands is a measure of the contributions 
of the respective orbitals. Note that these orbitals refer to the global 
coordinate system with the $ z $-axis along the chain direction. Obviously, 
all bands display strong dispersion parallel to the chain axis (i.e.\ 
the line $ \Gamma $-A), reflecting the one-dimensionality of the compounds, 
while weak dispersions perpendicular to this line indicate small 
interchain coupling. According to Figs.\ \ref{fig3} and \ref{fig4}, the most  
strongly dispersing bands are of almost pure $ d_{3z^2-r^2} $ character 
with similar contributions from both transition metal sites, pointing 
to the strong metal-metal bonding within the chains. While the band 
dispersions are quite similar for both cobaltates and the ferrate, 
important differences arise from the reduced band filling in the 
latter, which drastically alters the Fermi surface. In general, due to 
the almost perfect one-dimensional dispersion, the Fermi surfaces 
consist of flat sheets perpendicular to the $ c $ axis, i.e.\ the line 
$ \Gamma $-A. However, while in the cobaltates the $ z $-position of 
the Fermi surface is at the $ \Gamma $-point, we observe perfect 
nesting in $ {\rm Ca_3FeRhO_6} $ with the Fermi surface sheets at 
the A-point. As a consequence, the ferrate, but not the cobaltates, 
is susceptible to a Fermi surface instability against an antiferrodistortive 
mode or else the observed antiferromagnetic intrachain order.

\section{Conclusion}

As for the previously investigated $ {\rm Ca_3Co_2O_6} $, electronic 
structure calculations for $ {\rm Ca_3CoRhO_6} $ and $ {\rm Ca_3FeRhO_6} $ 
reveal strong effects of the local coordination on the electronic and 
magnetic properties. In particular, low- and high-spin moments are 
obtained at the octahedral Rh sites and at the trigonal prismatic 
Co/Fe sites, respectively. Strong $ d $--$ p $ hybridization results 
in considerable oxygen polarization, giving rise to the formation of 
extended magnetic moments localized at the prisms. Metal-metal bonding 
via Rh $ 4d_{3z^2-r^2} $ orbitals lays ground for large contributions 
to the intrachain exchange coupling. Differences between the cobaltates 
and the ferrate due to the reduced electron count in the latter show 
up in the additional polarization of the Fe $ 3d_{xy,x^2-y^2} $ states. 
Furthermore, the corresponding band shifts have drastic effects for 
the Fermi surface, leading to almost perfect nesting and an 
A-point instability, which drives the observed antiferromagnetic 
intrachain order.

\section*{Acknowledgements}
C.\ Hackenberger was supported by a Marie Curie fellowship of the
European Community program under number HPMT2000-141. This work was 
supported by the Deutsche Forschungsgemeinschaft (SFB 484) and by 
the BMBF (13N6918).


\begin{thebibliography}{00}

\bibitem{stitzer01}
K.\ E.\ Stitzer, J.\ Darriet, and H.-C.\ zur Loye,
Curr.\ Opin.\ Solid State Mater.\ Sci.\ {\bf 5}, 535 (2001).

\bibitem{fjellvag96}
H.\ Fjellv\aa g, E.\ Gulbrandsen, S.\ Aasland, A.\ Olsen, and B.\ C.\ Hauback,
J.\ Solid State Chem.\ {\bf 124}, 190 (1996).

\bibitem{aasland97}
S.\ Aasland, H.\ Fjellv\aa g, and B.\ C.\ Hauback,
Solid State Comm.\ {\bf 101}, 187 (1997).

\bibitem{kageyama97}
H.\ Kageyama, K.\ Yoshimura, K.\ Kosuge, H.\ Mitamura, and T.\ Goto,
J.\ Phys.\ Soc.\ Japan {\bf 66}, 1607 (1997).

\bibitem{hardy03a}
V.\ Hardy, S.\ Lambert, M.\ R.\ Lees, and D.\ McK.\ Paul, 
Phys.\ Rev.\ B {\bf 68}, 014424 (2003). 

\bibitem{sugiyama05}
J.\ Sugiyama, H.\ Nozaki, J.\ H.\ Brewer, E.\ J.\ Ansaldo, T.\ Takami, 
H.\ Ikuta, and U.\ Mizutani, 
Phys.\ Rev.\ B {\bf 72}, 064418 (2005). 

\bibitem{maignan00}
A.\ Maignan, C.\ Michel, A.\ C.\ Masset, C.\ Martin, and B.\ Raveau,
Eur.\ Phys.\ J.\ B {\bf 15}, 657 (2000).

\bibitem{hardy04}
V.\ Hardy, M.\ R.\ Lees, O.\ A.\ Petrenko, D.\ McK.\ Paul, D.\ Flahaut, 
S.\ H\'ebert, and A.\ Maignan, 
Phys.\ Rev.\ B {\bf 70}, 064424 (2004). 

\bibitem{eyert04}
V.\ Eyert, C.\ Laschinger, T.\ Kopp, and R.\ Fr\'esard,
Chem.\ Phys.\ Lett.\ {\bf 385}, 249 (2004).

\bibitem{laschinger04}
C.\ Laschinger, T.\ Kopp, V.\ Eyert, and R.\ Fr\'esard,
J.\ Magn.\ Magn.\ Mat.\ {\bf 272-276}, 974 (2004).

\bibitem{fresard04}
R.\ Fr\'esard, C.\ Laschinger, T.\ Kopp, and V.\ Eyert,
Phys.\ Rev.\ B {\bf 69}, 140405(R) (2004). 

\bibitem{dissudo}
U.\ Schwingenschl\"ogl, PhD thesis, Universit\"at Augsburg 2004.

\bibitem{niitaka99}
S.\ Niitaka, H.\ Kageyama, M.\ Kato, K.\ Yoshimura, and K.\ Kosuge, 
J.\ Solid State Chem.\ {\bf 146}, 137 (1999).

\bibitem{niitaka01}
S.\ Niitaka, K.\ Yoshimura, K.\ Kosuge, M.\ Nishi, and K.\ Kakurai, 
Phys.\ Rev.\ Lett.\ {\bf 87}, 177202 (2001).

\bibitem{niitaka02}
S.\ Niitaka, K.\ Yoshimura, K.\ Kosuge, A.\ Mitsuda, H.\ Mitamura, 
and T.\ Goto,
J.\ Phys.\ Chem.\ Solids {\bf 63}, 999 (2002).

\bibitem{davis03}
M.\ J.\ Davis, M.\ D.\ Smith, and H.-C.\ zur Loye,
J.\ Solid State Chem.\ {\bf 173}, 122 (2003).

\bibitem{hardy03b}
V.\ Hardy, M.\ R.\ Lees, A.\ Maignan, S.\ H\'ebert, D.\ Flahaut, and 
D.\ McK.\ Paul, 
J.\ Phys.: Condens.\ Matt.\ {\bf 15}, 5737 (2003). 

\bibitem{niitaka01b}
S.\ Niitaka, H.\ Kageyama, K.\ Yoshimura, K.\ Kosuge, S.\ Kawano, 
N.\ Aso, A. Mitsuda, H.\ Mitamura, and T.\ Goto,
J.\ Phys.\ Soc.\ Jpn.\ {\bf 70}, 1222 (2001).

\bibitem{loewenhaupt03}
M.\ Loewenhaupt, W.\ Sch\"afer, A.\ Niazi, and E.\ V.\ Sampathkumaran,
Europhys.\ Lett.\ {\bf 63}, 374 (2003).

\bibitem{niitaka03}
S.\ Niitaka, K.\ Yoshimura, K.\ Kosuge, K.\ Mibu, H.\ Mitamura, and T.\ Goto, 
J.\ Magn.\ Magn.\ Mat.\ {\bf 260}, 48 (2003).

\bibitem{wkg} 
A.\ R.\ Williams, J.\ K\"ubler, and C.\ D.\ Gelatt jr.,
Phys.\ Rev.\ B {\bf 19}, 6094 (1979).

\bibitem{revasw} 
V.\ Eyert, 
Int.\ J.\ Quant.\ Chem.\ {\bf 77}, 1007 (2000).

\bibitem{eyert98b}
V.\ Eyert and K.-H.\ H\"ock,
Phys.\ Rev.\ B {\bf 57}, 12727 (1998).

\bibitem{mixpap}
V.\ Eyert,
J.\ Comp.\ Phys.\ {\bf 124}, 271 (1996).






\end{thebibliography}
\end{document}